\documentclass[prb,twocolumn,superscriptaddress,showpacs]{revtex4}

\usepackage{amsmath}
\usepackage{amssymb}
\usepackage{epsfig}
\usepackage{graphicx}
\usepackage{wasysym}
\usepackage{bbm}

\newcommand \be{\begin{equation}}
\newcommand \ee{\end{equation}}
\newcommand \bes{\begin{equation*}} 
\newcommand \ees{\end{equation*}}
\newcommand \bea{\begin{eqnarray}}
\newcommand \eea{\end{eqnarray}}
\newcommand \bsea{\begin{subequations}\begin{eqnarray}} 
\newcommand \esea{\end{eqnarray}\end{subequations}}
\newcommand \beas{\begin{eqnarray*}} 
\newcommand \eeas{\end{eqnarray*}}
\newcommand \bfg{\begin{figure}}
\newcommand \efg{\end{figure}}
\newcommand \bfgs{\begin{figure*}} 
\newcommand \efgs{\end{figure*}}
\newcommand \bwt{\begin{widetext}}
\newcommand \ewt{\end{widetext}}
\newcommand \nd{{\vphantom{\dagger}}} 

\newcommand \im{\mathbbm{i}} 

\newcommand \ve{\mathbf{e}}
\newcommand \vE{\mathbf{E}}
\newcommand \vK{\mathbf{K}}
\newcommand \vk{\mathbf{k}}
\newcommand \vB{\mathbf{B}}
\newcommand \vD{\mathbf{D}}
\newcommand \vS{\mathbf{S}}
\newcommand \vecr{\mathbf{r}}
\newcommand \vsigma{\boldsymbol{\sigma}}
\newcommand \sublat{\mathcal{V}}
\newcommand \edif{\epsilon} 

\begin{document}

\title{Nearly flat band with Chern number $C=2$ on the dice lattice}


\author{Fa Wang}
\affiliation{Department of Physics, Massachusetts Institute of Technology, 
Cambridge, Massachusetts 02139, USA}

\author{Ying Ran}
\affiliation{Department of Physics, Boston College, 
Chestnut Hill, Massachusetts 02467, USA}

\date{\today}

\begin{abstract}
We point out the possibility of a nearly flat band with Chern number $C=2$ 
on the dice lattice in a simple nearest-neighbor tightbinding model. 
This lattice can be naturally formed by
three adjacent $(111)$ layers of cubic lattice, 
which may be realized in certain thin films or artificial heterostructures, 
such as the SrTiO$_3$/SrIrO$_3$/SrTiO$_3$ trilayer heterostructure grown along 
the $(111)$ direction.
The flatness of two bands is protected by the bipartite nature 
of the lattice.
Including the Rashba spin-orbit coupling on nearest-neighbor bonds 
separate the flat bands from the others but maintains their flatness.
Repulsive interaction will drive spontaneous ferromagnetism 
on the Kramer pair of the flat bands
and split them into two nearly flat bands with Chern number $C=\pm 2$.
We thus propose that this may be a route to the quantum anomalous Hall effect 
and further conjecture that the partial filling of the $C=2$ band
may realize exotic fractional quantum Hall effects.
\end{abstract}

\pacs{71.10.Fd,73.43.Cd,73.20.At}

\maketitle

A few years after the experimental discovery of integer quantum hall effect(IQHE)\cite{Klitzing}, Haldane wrote down a tight-binding model on the honeycomb lattice with IQHE\cite{Haldane88}, explicitly showing that the essence of IQHE is \emph{not} the external magnetic field. However, it takes more than two decades for people to show that the similar statement is also true for fractional quantum Hall effect(FQHE).
Recently several groups have proposed to realize 
FQHE without Landau levels\cite{Wen, Sun, Chamon,ShengDN,Fiete,Bernevig}.
The basic idea is to engineer a nearly flat band in two dimensions(2D)
with nonzero Chern number. 
Electron interaction in this partially filled band may realize 
fractional quantum Hall effect, as suggested by
exact diagonalization studies\cite{Sun,Chamon,ShengDN,Bernevig}.

\begin{figure}
\includegraphics[scale=0.4]{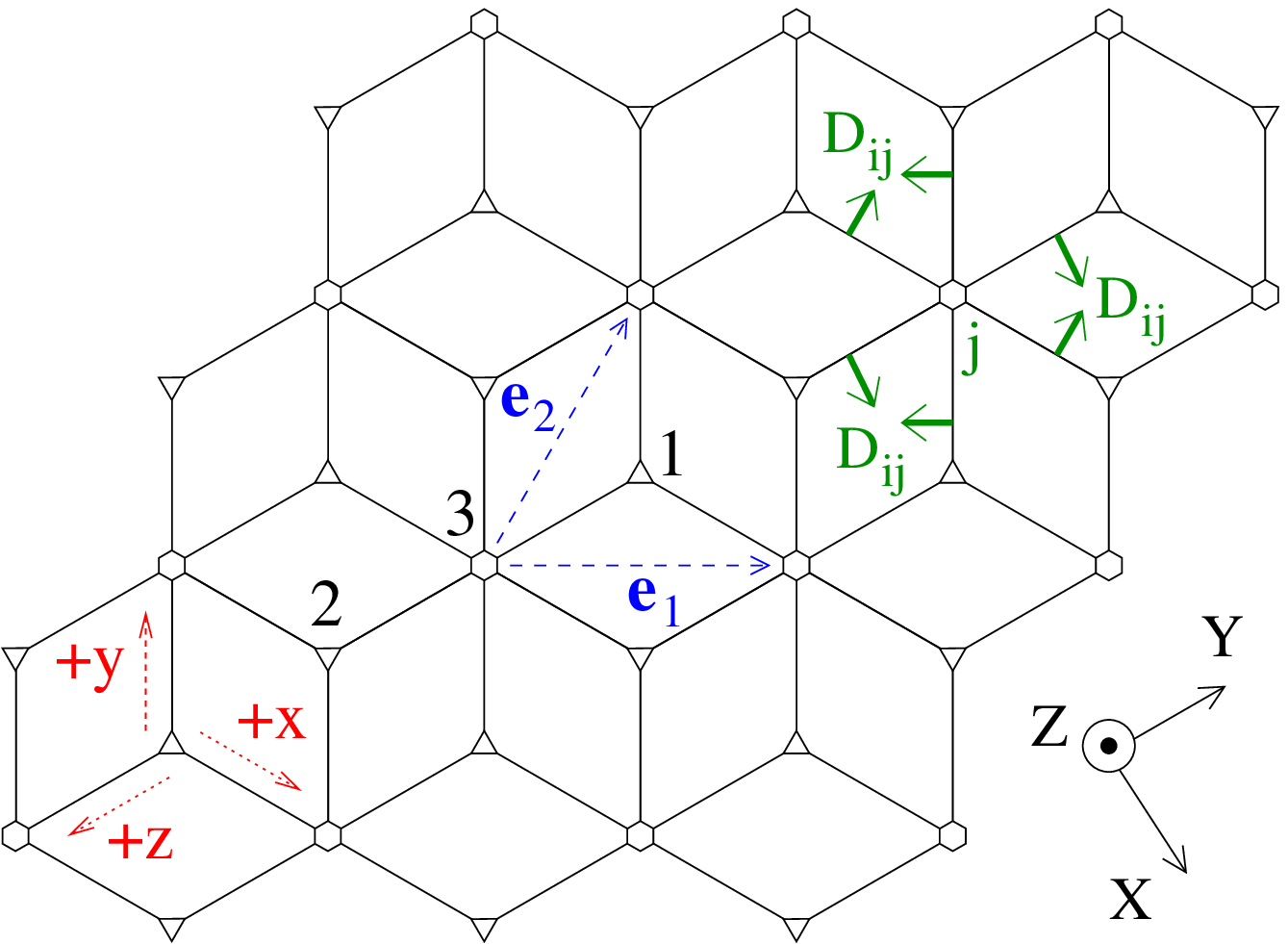}
\includegraphics[scale=0.5]{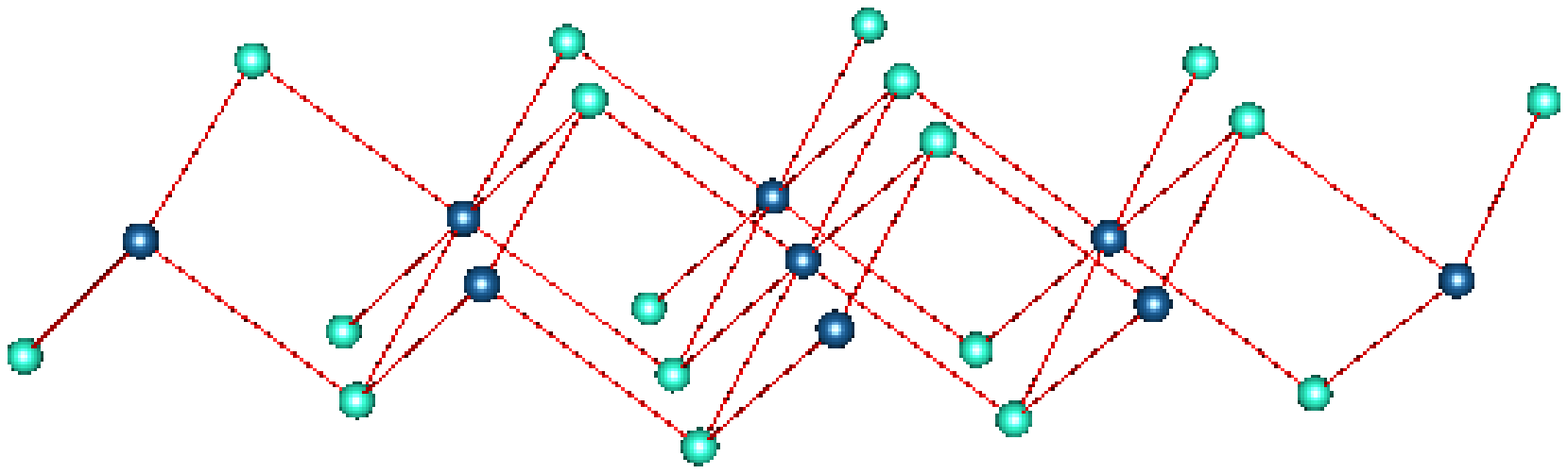}
\caption{(Color online) 
Top: The dice lattice. 
Small upward triangles (bottom layer), downward triangles (top layer), 
and hexagons (middle layer) indicate the three sublattices.
$1,2,3$ label the three basis sites in the unit cell at origin.
Coordination-number-3 sites ($1$ and $2$) and
coordination-number-6 sites ($3$) are
the two subsets of this bipartite lattice.
Blue dash arrows labelled by $\ve_1$ and $\ve_2$ indicate 
the two translations of the dice lattice.
Thick green arrows labelled as $D_{ij}$ indicate
the Rashba SOC directions on those bonds $ij$, 
with coordination-number-6 site $j$.
Red dotted arrows with the labels $+x,+y,+z$ indicate of the
projection of cubic lattice axis.
Captial $X,Y,Z$ are axis for spin space in Rashba SOC.
$Z$ is the original $(111)$ direction.
Bottom: Perspective view of three adjacent $(111)$ layers of cubic lattice. 
The middle layer has different color for easy recognition.
The top view of this tri-layer is the dice lattice.
}
\label{fig:dice}
\end{figure}

In these proposals nearly flat bands are obtained by
fine-tuning ratios between nearest-neighbor(NN), next-nearest-neighbor(NNN),
and even further neighbor tightbinding parameters. 
In this paper
we point out a route to get {\em completely} flat bands 
without this fine-tuning by employing a bipartite lattice 
with unequal number of two subsets of sites\cite{Lieb}.
As a concrete example we consider the dice lattice as shown in 
Fig.~\ref{fig:dice}. 
It is bipartite with unequal number of two subsets of sites 
(the coordination-number-3 sites are twice as many as 
the coordination-number-6 sites). This system is inversion symmetric with respect to the coordination-number-6 sites. 
We consider a single $s$-orbital with spin-1/2 degrees of freedom(DOF) on every site, and
mainly focus on systems close to half-filling, i.e., one electron per site. The NN tight-binding model, including 
the Rashba-type spin-orbit coupling(SOC) consistent with lattice symmetry,
will produce two completely flat bands separated from the 
other bands. Because the two flat bands are half-filled, ferromagnetism is a natural consequence of 
correlation\cite{Lieb,Mielke,Tasaki,Stoner},
which gives rise to a Zeeman field on the mean-field level. We demonstrate the spontaneous ferromagnetism by a variational wave function study of Hubbard interactions.

As a nice feature of the current model system, even a small Zeeman field can split this Kramer pair of flat bands
and produce two \emph{separated} nearly flat bands with Chern number $C=\pm 2$.
Filling one of them will then produce quantized anomalous Hall(QAH) effect with $\sigma_{xy}=2\frac{e^2}{h}$. This Zeeman field could also be extrinsic, e.g. growing the system on a ferromagnetic substrate. Note that in a usual ferromagnetic system, a realistic Zeeman splitting would not completely separate the two bands with opposite spin polarizations, and a ferromagnetic \emph{metal} results. This is partially why the QAH insulator, which needs to be a ferromagnetic insulator, has not been realized experimentally so far. The main advantage of the presently studied system is the existence of the half-filled flat bands, which natually support well-separated bands by a realistic Zeeman splitting.

\emph{Material realizations:} 
This model hamiltonian may actually be relevant to some real systems. 
Heterostructures of transition metal oxide(TMO) perovskites, whose crystal structures are cubic, are becoming available owing to the recent development \cite{Izumi200153,Ohtomo2002,Ohtomo2004} in the fields of oxide superlattices and oxide electronics(for a review, see \cite{Mannhart26032010}). In particular, layered structures of TMO heterstructures can now be prepared with atomic precision, thus offering a high degree of control over important material properties, such as lattice constant, carrier concentration, spin-orbit coupling, and correlation strength. 

TMO heterostructures grown along the $(111)$ direction have been synthesized experimentally (e.g., Ref.~\cite{10.1063/1.3525578,10.1063/1.3455323}).  Recently it was pointed out that TMO (111) bilayer heterostructures are promising candidates hosting various topological phases of matter\cite{Xiao2011}. The dice lattice here can be formed by three adjacent $(111)$ layers of cubic lattice, each of which is a triangular lattice (Fig.~\ref{fig:dice}).
Although we considered only simple $s$-orbital on every site here,
the result should be valid if the active orbital is a one dimensional 
representation of the $D_{3d}$ group.
Some examples are the $p_Z$-orbital ($p_x+p_y+p_z$),
and the $a_{\rm 1g}$ orbital ($d_{yz}+d_{zx}+d_{xy}$) of $d$-electrons under 
cubic and trigonal crystal potential.

A particularly relevant example is the transition metal oxide SrTiO$_3$/SrIrO$_3$/SrTiO$_3$ trilayer heterostructure. 
Note that although the crystal structure of the bulk SrIrO$_3$ is a monoclinic distortion
of the hexagonal BaTiO$_3$ structure\cite{Longo1971174}, thin films of perovskite SrIrO$_3$ have been synthesized on substrates\cite{perovskite_SrIrO3}, which are reported to be metallic\cite{Kim}. This indicates that an itinerant electronic model could be a good starting point of describing the SrTiO$_3$/SrIrO$_3$/SrTiO$_3$ trilayer heterostructure. Due to the strong spin-orbit coupling on the Ir$^{4+}$ ion, together with the octahedral crystal field, the active orbital is a half-filled effective $J_{\rm eff}=1/2$ doublet\cite{Kim}. The explictly form of these doublet in the presence of cubic symmetry is $
 |J_z=1/2\rangle=\frac{1}{\sqrt{3}} (+i|xy,\uparrow\rangle-|xz,\downarrow\rangle+i|yz,\downarrow\rangle)$, and $|J_z=-1/2\rangle$ is its time-reversal partner. 
These half-filled orbitals hop around the dice lattice, and contribute to states close to the fermi level.  indicating the correlation in the bulk system is intermediate. In a (111) heterostructure, cubic symmetry is reduced to trigonal symmetry. Nevertheless, to the leading order with respect to trigonal distortion, the nearest neighbor hoppings between these $J=1/2$ orbitals are identical to the hoppings of the $s$-orbitals, which form the model hamiltonian considered in this paper.

Therefore we think that our proposal is a promising route to realize QAH effect. In the same spirit of previous works on the FQHE without Landau levels, 
we conjecture that fractional filling of these bands might produce exotic fractional quantum hall (FQH) states. The nature of these FQH states remains unclear and we leave it as a subject of future research. But it is worth pointing out that, in a nearly flat band with Chern number $C=2$, the natural candidate ground states for $\nu=1/m$ ($m$ is odd integer) filling fractions are non-abelian states described by $SU(m)_2$ Chern-Simons effective theory\cite{Lu_Ran}.

{\em NN model without SOC.} 
The dice lattice and coordination system are defined in Fig.~\ref{fig:dice}. 
Label the three sublattices by $\sublat_{1,2,3}$ respectively. 
Consider a single $s$-orbital with spin-1/2 DOF on every site. 
As a warmup consider NN spin-independent hopping only,
\be
H_0=-\sum_{<ij>,\alpha}(t\,c_{i\alpha}^\dagger c_{j\alpha}^\nd+{h.c.})
-\sum_{i\in\sublat_3}\edif\,n_i,
\ee
where $\alpha=\uparrow,\downarrow$ labels spin, $i,j$ label sites, 
$n_i=\sum_{\alpha}c_{i\alpha}^\dagger c_{i\alpha}^\nd$ is 
the electron density on site $i$. 
Note that sublattice-$3$ ($\sublat_3$) has 
a onsite energy difference $\edif$ from the other two sublattices, 
as is allowed by symmetry.
In this section the spin DOF will be omitted.

\begin{figure}
\includegraphics[scale=0.65]{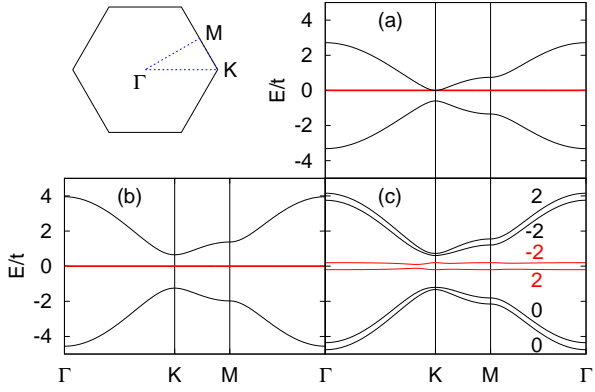}
\caption{(Color online).
Dispersions of NN tightbinding models on dice lattice
along high symmetry directions. 
Parameters used are $\edif=0.6\,t$, $\lambda=0.3\,t$,
and $B=0.2\,t$. 
The bands marked as red in the middle are the (nearly) flat bands.
Top-left corner is the Brillouin zone with the 
high symmetry lines indicated by dash blue lines.
(a). Spin-independent hoppings only.
(b). Spin-independent hoppings plus Rashba SOC $\lambda$.
(c). Spin-independent hoppings plus Rashba SOC $\lambda$ 
and magnetic field $B$ along $Z$ ($111$) direction.
The Chern number of each band is indicated.
}
\label{fig:dispersion}
\end{figure}

In momentum space the hamiltonian reads
\be
H_0(\vk)=-\begin{pmatrix}
0 & 0 & t\,\gamma_{\vk}^* \\
0 & 0 & t\,\gamma_{\vk}^\nd \\
t\,\gamma_{\vk}^\nd & t\,\gamma_{\vk}^* & \edif
\end{pmatrix},
\ee
where $\gamma_{\vk}=1+e^{\im k_1}+e^{\im k_2}$, 
$\im=\sqrt{-1}$,  
$k_{1,2}=\vk\cdot \ve_{1,2}$ respectively,
and the basis is $(c_{1\vk},c_{2\vk},c_{3\vk})$. 
This model has three bands with dispersions
$E_1=-\edif/2-\sqrt{\edif^2/4+2\,t^2|\gamma_{\vk}|^2}$,
$E_2=0$, and 
$E_3=-\edif/2+\sqrt{\edif^2/4+2\,t^2|\gamma_{\vk}|^2}$,
as illustrated in Fig.~\ref{fig:dispersion}(a). 

The middle band is completely flat as required by the bipartiteness. 
However the top band touches the flat bands quadratically 
at the Brillouin zone corners $\pm \vK=\pm (k_1=4\pi/3,k_2=2\pi/3)$,
similar to double-layer graphene\cite{bilayer-graphene}
or certain other models with flat bands\cite{Bergman}. 
The effective two-band hamiltonian at 
the band touching point $\pm\vK$ is
\be
\frac{3\,t^2}{4\,\edif}\begin{pmatrix}
|\delta\vk|^2 & \delta k_{\pm}^2\\
\delta k_{\mp}^2 & |\delta\vk|^2
\end{pmatrix}+O(\delta\vk^4),
\ee
where $\delta\vk=\vk\mp\vK$, 
$\delta k_{+}=e^{\im \pi/3}(\delta k_X+ \im \delta k_Y)$
and $\delta k_{-}=(\delta k_{+})^*$. 

The flat band has Bloch wavefunction $(\gamma_{\vk}^*,-\gamma_{\vk}^\nd,0)$ 
on the three sublattices.
It has local Wannier functions
residing on the six neighbors of a coordination-number-6 site (sublattice-$3$)
with opposite amplitudes between sublattice-$1$ and sublattice-$2$.

{\em NN model with Rashba SOC.}
Rashba SOC induced by electric fields can be included as
\be
H_{0,{\rm SOC}}=H_0
-\sum_{<ij>,\alpha,\beta}[\im\lambda\,
c_{i\alpha}^\dagger 
(\hat{\vD}_{ij}\cdot\vsigma)_{\alpha\beta}^\nd
c_{j\beta}^\nd
+{h.c.}
],
\ee
where $\vsigma$ are spin Pauli matrices, 
$\hat{\vD}_{ij}$ is the unit vector along the direction
of the cross product $\vE_{ij}\times \vecr_{ij}$ 
of electric field $\vE_{ij}$ and displacement $\vecr_{ij}$ 
for bond $ij$, 
$\lambda$ is the strength of SOC and is uniform 
on all NN bonds as required by translation and 
$D_{3d}$ symmetry of the tri-layer. 
The $D_{3d}$ symmetry further restricts the direction of $\vE_{ij}$ 
to be within the vertical [perpendicular to the $(111)$ layers] 
plane containing bond $ij$. 
Therefore $\hat{\vD}_{ij}$ are all parallel to the $(111)$ layers. 
Their directions are illustrated in Fig.~\ref{fig:dice}.
In momentum space the hamiltonian reads
\be
\begin{split}
& H_{0,{\rm SOC}}(\vk)=
H_0(\vk)\otimes\mathbf{1}_{2\times 2}-\im\lambda
\\ &
\times
\begin{pmatrix}
0 & 0 & 0 & 0 & 0 & \gamma_{\vk+}^*\\
0 & 0 & 0 & 0 & \gamma_{\vk-}^* & 0\\
0 & 0 & 0 & 0 & 0 & \gamma_{\vk-}^\nd\\
0 & 0 & 0 & 0 & \gamma_{\vk+}^\nd & 0\\
0 & -\gamma_{\vk-}^\nd & 0 & -\gamma_{\vk+}^* & 0 & 0\\
-\gamma_{\vk+}^\nd & 0 & -\gamma_{\vk-}^* & 0 & 0 & 0\\
\end{pmatrix},
\end{split}
\ee
where $\gamma_{\vk\pm}=1+e^{\im (k_1\pm 2\pi/3)}+e^{\im (k_2\pm 4\pi/3)}$,
and the basis is $(
c_{1\vk\uparrow},c_{1\vk\downarrow},
c_{2\vk\uparrow},c_{2\vk\downarrow},
c_{3\vk\uparrow},c_{3\vk\downarrow}
)$. 
It has three doubly degenerate bands with dispersions
$E_1=-\edif/2-\sqrt{\edif^2/4+2\,t^2|\gamma_{\vk}|^2+\lambda^2 (|\gamma_{\vk,-}|^2+|\gamma_{\vk,+}|^2)}$, 
$E_2=0$,
$E_3=-\edif/2+\sqrt{\edif^2/4+2\,t^2|\gamma_{\vk}|^2+\lambda^2 (|\gamma_{\vk,-}|^2+|\gamma_{\vk,+}|^2)}$.
If $\lambda \ll \edif$, the effective four-band hamiltonian
at the original band touching point $\pm K$ is
\be
\begin{split}
&
\frac{3\,t^2}{4\,\edif}\begin{pmatrix}
|\delta\vk|^2 & \delta k_{\pm}^2\\
\delta k_{\mp}^2 & |\delta\vk|^2
\end{pmatrix}\otimes \mathbf{1}_{2\times 2}
+
\frac{3\sqrt{3}\,\lambda\,t}{2\,\edif}\mathbf{1}_{2\times 2}\otimes
\begin{pmatrix}
0 & k_{\mp}\\
k_{\pm} & 0
\end{pmatrix}
\\ &
+
\frac{9\lambda^2}{2\,\edif}
\left [\mathbf{1}_{4\times 4}
\pm 
\tau^z
\otimes \sigma^z\right ]
+O(\lambda^2\delta\vk)+O(\delta\vk^4),
\end{split}
\ee
where Pauli matrix $\tau^z$ acts on the sublattices-$1,2$ space.
The mass term $\tau^z\otimes \sigma^z$ has opposite sign between
the two band touching points $\pm K$, 
similar to the Haldane model\cite{Haldane88}. 
The dispersions are illustrated in Fig.~\ref{fig:dispersion}(b).
There are still two completely flat band dictated by
the hamiltonian structure.

\begin{figure}
\includegraphics[scale=0.6]{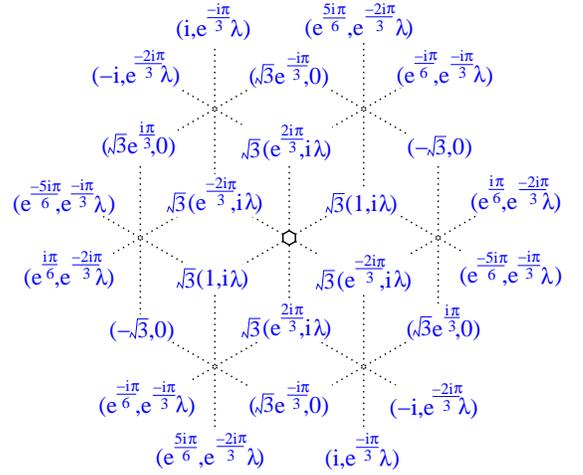}
\caption{(Color online) 
Local Wannier function of one of the flat bands 
of NN model on dice lattice (black dotted lines) with Rashba SOC $\lambda$
($t=1$ for simplicity). 
The two component vector (blue) on each coordination-number-3
site indicates spin-up and spin-down amplitudes on that site. 
Amplitudes on coordination-number-6 sites vanish,
therefore the parameter $\edif$ has no effect. 
The small hexagon is the ``guiding center'' of this Wannier function.
}
\label{fig:Wannier}
\end{figure}

The flat bands have Bloch wavefunctions
$(
t\,\gamma_{\vk}^* \Gamma_{\vk}^\nd,
\im\lambda\,\gamma_{\vk+}^* \Gamma_{\vk}^*,
-t\,\gamma_{\vk}^\nd \Gamma_{\vk}^\nd,
-\im\lambda\,\gamma_{\vk-}^\nd \Gamma_{\vk}^*,
0,0)$
and its Kramer pair,
where $\Gamma_{\vk}^\nd=
\gamma_{\vk}^\nd \gamma_{\vk+}^* - \gamma_{\vk}^*\gamma_{\vk-}^\nd$.
Therefore these flat bands also have local Wannier functions.
One of the Wannier functions is illustrated in Fig.~\ref{fig:Wannier}.
The other Wannier functions can be produced by translation and 
time-reversal. 
Note that the spin-up component of 
the illustrated Wannier function acquires a phase $2\pi/3$ 
under six-fold rotation around its ``guiding center'', 
similar to the cyclotron orbit in Landau levels
except that the phase here is twice as large, 
suggesting Chern number $C=2$.

{\em Nearly flat band with Chern number $C=2$.}
The previous double-degeneracy of the flat bands is 
protected by time-reversal symmetry. 
Consider magnetic field effect,
\be
H_{0,{\rm SOC}+B}
=H_{0,{\rm SOC}}
-g\sum_{i}\vB_{i}\cdot\vS_i
\ee
where $\vS_i=(1/2)\sum_{\alpha,\beta}
c_{i\alpha}^\dagger \vsigma_{\alpha\beta}^\nd c_{i\beta}^\nd$
is electron spin,
the Bohr magneton $\mu_B$ is omitted, 
and $g=2$ 
($g=-2$ for the $J_{\rm eff}=1/2$ states of Ir$^{4+}$ ion\cite{Bleaney})
is assumed hereafter.
The field on sublattice-$1,2$ may be different
from that on sublattice-$3$. 
For illustration purpose we draw the band structure 
with uniform field $B_{i}=0.2\,t$ along $(111)$ direction ($Z$ direction)
in Fig.~\ref{fig:dispersion}(c).
As expected the Kramer pair of flat bands split into two
nearly flat bands. 
Direct computation of Chern numbers shows that they
carry Chern number $C=\pm 2$ [Fig.~\ref{fig:dispersion}(c)].

There is a simple physical argument that proves the $C=\pm 2$ for the two nearly flat bands. Let us turn off the Rashba-type spin-orbit coupling $\lambda$ for the moment. Note that the sublattice-1,2 form a honeycomb lattice by themselves. We could turn on another artificial 
$i\lambda_1\vsigma\cdot (1,1,1)$ spin-orbit coupling between the second-neighbors on these two sublattices only, with the same signs as the Kane-Mele model\cite{KaneMele}. In this $\lambda_1$-only model, spin rotation along (111) direction is conserved so that we could consider each spin-poloarized subsystem separately. Energy gaps at $K$ and $-K$, the quadratic band-touching points, in Fig.~\ref{fig:dispersion}(a) are opened by $\lambda_1$. However, it is well-known that the spin-orbit energy gap at a quadratic band-touching point transfers Chern number one between the two bands. Therefore, two spin-orbit gaps at $K$ and $-K$ transfer Chern number \emph{two} instead of one as in the Kane-Mele case. Including a Zeeman field $\vB$ along (111) split the $C=\pm2$ bands, and clearly, the resulting bands and their Chern numbers must be very similar to those of the $\lambda$-only model [see Fig.~\ref{fig:dispersion}(c)]. Fixing a Zeeman field $\vB$, it turns out that one can adiabatically connect the $\lambda_1$-only model with the $\lambda$-only model by interpolation while keeping all the six bands isolated from one another. This adiabatic evolution perserves the Chern numbers of each bands. We thus prove the Chern numbers in Fig.~\ref{fig:dispersion}(c).

{\em Spontaneous ferromagnetism.} 
The flat band is half-filled if the entire system is at half-filling.
Add onsite Hubbard interactions in the hamiltonian,
\be
H_{\rm int}=H_{0,{\rm SOC}}+\sum_{i} U\, n_{i\uparrow}n_{i\downarrow}.
\ee
If SOC $\lambda=0$, by Lieb's theorem\cite{Lieb} the ground state is 
ferromagnetic with total spin $S=(1/2)[(N_1+N_2)-N_3]=(1/2)N_{\rm cell}$ 
($N_{1,2,3}$ is the number of sites on sublattice-$1,2,3$ respectively, 
and equals to the number of unit cells $N_{\rm cell}$). 
With Rashba $\lambda$ there is no known proof of ferromagnetism. 
We use a variational (mean field) treatment of this problem. 

\begin{figure}
\includegraphics[scale=0.52]{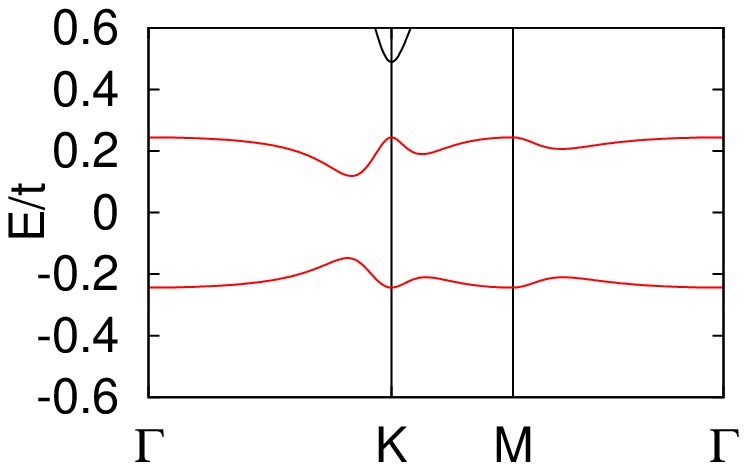}
\includegraphics[scale=0.52]{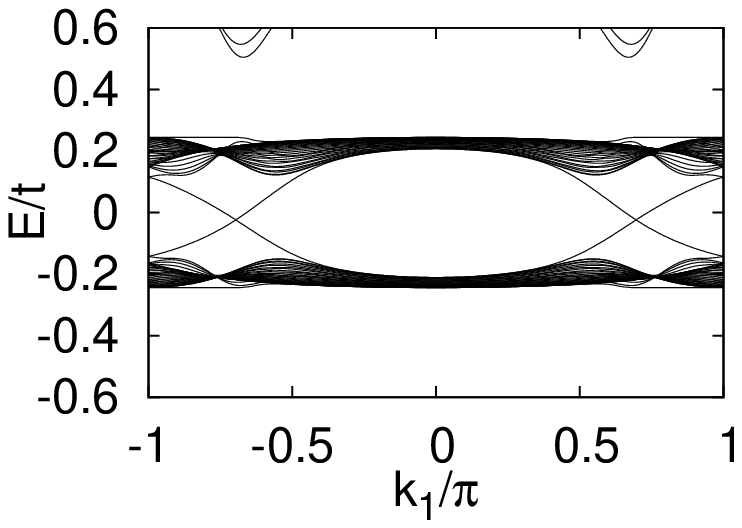}
\caption{(Color online) 
Left: 
The two nearly flat bands (red) with Chern numbers $C=\pm 2$. 
Parameters are $\edif=0.6\,t$, $\lambda=0.3\,t$, 
$B_1=0.2440\,t$, and $B_3=-0.0162\,t$. 
Right: 
Dispersion of a cylinder with 32 unit cell 
open boundary condition along $\ve_2$
and periodic boundary condition along $\ve_1$,
showing the edge states between the nearly flat bands.
}
\label{fig:detail-dispersion}
\end{figure}

The ferromagnetic ``mean field'' hamiltonian is
just the free fermion hamiltonian with magnetic field 
$H_{0,{\rm SOC}+B}$. 
By inversion symmetry we assume field on sublattices-$1,2$
are the same, $\vB_2=\vB_1$, but may be different from 
that on sublattice-$3$, $\vB_3$. 
The variational wavefunction is the free fermion wavefunction 
by half-filling this mean field hamiltonian. 
We then evaluate the energy expectation value of the Hubbard model
$H_{\rm int}$ and try to minimize it with respect to
the variational parameters $\vB_1$ and $\vB_3$.
From preliminary numerical results,
the system is unstable to spontaneous ferromagnetism for infinitesmal 
repulsive $U$, 
consistent with the Stoner criterion\cite{Stoner},
however the energy gain is very insensitive to the field directions. 
For $\edif=0.6\,t$, $\lambda=0.3\,t$, $U=t$, 
and the field directions along $(111)$ ($Z$ direction),
the field strength is $B_1=0.2440\,t$ on sublattices-$1,2$ 
and $B_3=-0.0162\,t$ on sublattice-$3$. 
The mean field band structure is very similar to Fig.~\ref{fig:dispersion}(c) 
where a uniform $B=0.2\,t$ is used. 
The two nearly flat mean field bands are drawn in 
Fig.~\ref{fig:detail-dispersion}.
The three occupied mean field bands have total Chern number $C=2$ 
and exhibit anomalous quantum Hall effect.
The edge state on a cylindrical geometry is also shown in
Fig.~\ref{fig:detail-dispersion}.

{\em Conclusion.}
In this paper we discuss a model with spin-orbit coupling on the dice-lattice and the correlation physics in it. A transition metal oxide SrTiO3/SrIrO3/SrTiO3 trilayer heterostructure grown along the (111) direction, where this model may be realized, is proposed. In this system, two degenerate flat bands at half-filling are found. Stoner's instability naturally leads to ferromagnetism and split the two bands, which gives rise two nearly flat bands with Chern number $\pm 2$. This indicate a promising route to realize QAHE. We further speculate that further doping into the nearly flat Chern bands could lead to FQHE without an external magnetic field. We hope these results could encourage experimental syntheses and characterization of the material proposed here, as well as future theoretical investigations on the nature of the possible FQH states.

FW thanks the Institute for Advanced Study at Tsinghua University 
for hospitality where part of this work was finished. 
YR is supported by the startup fund at Boston College.

\end{document}